\def\year{2021}\relax
\documentclass[letterpaper]{article} 
\usepackage{aaai21}  
\usepackage{times}  
\usepackage{helvet} 
\usepackage{courier}  
\usepackage[hyphens]{url}  
\usepackage{graphicx} 
\usepackage{natbib}  
\usepackage{caption} 
\usepackage{tablefootnote}
\usepackage[labelformat=simple]{subcaption}
\usepackage{booktabs}
\usepackage{multirow}
\usepackage{amsmath}
\usepackage{amsfonts}

\newcommand{\eg}{e.g., }
\newcommand{\ie}{i.e., }
\newcommand{\vpara}[1]{\vspace{0.05in}\noindent\textbf{#1 }}

\newcommand{\figref}[1]{Fig.~\ref{#1}}
\newcommand{\tableref}[1]{Table~\ref{#1}} 
\newcommand{\citea}[1]{\citeauthor{#1}~\shortcite{#1}}
\newcommand{\citepp}{~\citep}
\newcommand{\shuidichou}{Waterdrop Fundraising}

\title{How Medical Crowdfunding Helps People? A Large-scale Case Study on Waterdrop Fundraising }
\author{
    Junjie Huang$^{1,2}$, Huawei Shen$^{1}$, Qi Cao$^{1,2}$, Li Cai$^{3}$, Xueqi Cheng$^{1}$\\
 }
\affiliations{
    $^{1}$CAS Key Laboratory of Network Data Science and Technology, Institute of Computing Technology, \\Chinese Academy of Sciences,  Beijing, China\\
$^{2}$University of Chinese Academy of Sciences, Beijing, China\\
$^{3}$ Beijing Zongqing Xiangqian Science and Technology Co Ltd, Beijing, China \\
\{huangjunjie17s, shenhuawei, caoqi, cxq\}@ict.ac.cn,  caili@shuidihuzhu.com

}

\begin{document}

\maketitle

\begin{abstract}
While online medical crowdfunding achieved tremendous success, quantitative study about whether and how medical crowdfunding helps people remains little explored.
In this paper, we empirically study how online medical crowdfunding helps people using more than 27,000 fundraising cases in \shuidichou, one of the most popular online medical crowdfunding platforms in China.
We find that the amount of money obtained by fundraisers is broadly distributed, \ie a majority of lowly donated cases coexist with a handful of very successful cases.
We further investigate the factors that potentially correlate with the success of medical fundraising cases.
Profile information of fundraising cases, \eg geographic information of fundraisers, affects the donated amounts, since detailed description may increase the credibility of a fundraising case. One prominent finding lies in the effect of social network on the success of fundraising cases: the spread of fundraising information along social network is a key factor of fundraising success, and the social capital of fundraisers play an important role in fundraising.
Finally, we conduct prediction of donations using machine learning models, verifying the effect of potential factors to the success of medical crowdfunding.
Altogether, this work presents a data-driven view of medical fundraising on the web and opens a door to understanding medical crowdfunding.

\end{abstract}

\section{Introduction}
ArtistShare\footnote{https://www.artistshare.com/} launched in 2003 as the Internet’s first fan-funding (referred to today as ``crowdfunding'') platform for creative artists.
After that, lots of online crowdfunding platforms have emerged (\eg kickstarter\footnote{https://www.kickstarter.com/}) and play a role of intermediates to bridge the fundraisers and backers.
Different from commercial crowdfunding, medical crowdfunding is a way of fundraising for medical bills.
It has become a popular choice for people with unaffordable health needs, especially for low-income and middle-income certain families with limited social welfare.
Online platforms  for medical crowdfunding are rapidly emerging in many countries.
Several popular online medical crowdfunding platforms include GoFundMe\footnote{https://www.gofundme.com/} in the USA and Waterdrop Fundraising\footnote{https://www.shuidichou.com/} in China.
They all achieved immense success in the medical crowdfunding.
Despite the proliferation of online crowdfunding, little is known about whether and how medical crowdfunding helps people, especially lacking quantitative study.
Compared with GoFundMe in the USA which allows people to raise money for other life events (\eg celebrations and graduations), Waterdrop Fundraising is focused on medical crowdfunding.
Moreover, Waterdrop Fundraising is conducted on social platforms in China (\eg WeChat).
Therefore, spreading relevant fundraising information in social networks is an important factor for the success of crowdfunding campaigns.
The relationships among users of social platforms are relatively intimate social relationships (\eg family, relatives, and friends).
Based on these social platforms, Waterdrop Fundraising gives some guides for fundraisers and has developed a convenient online payment system, making donations and sharing easier.
These differences between GoFundMe in the USA and Waterdrop Fundraising in China raise new research demand.
In other words, for this kind of newly emerging medical crowdfunding platform, \ie Waterdrop Fundraising, quantitative researches on them will help to explore the trends and characteristics of online medical crowdfunding.

\begin{figure}[t]
  \centering   
  \includegraphics[width=0.45\textwidth]{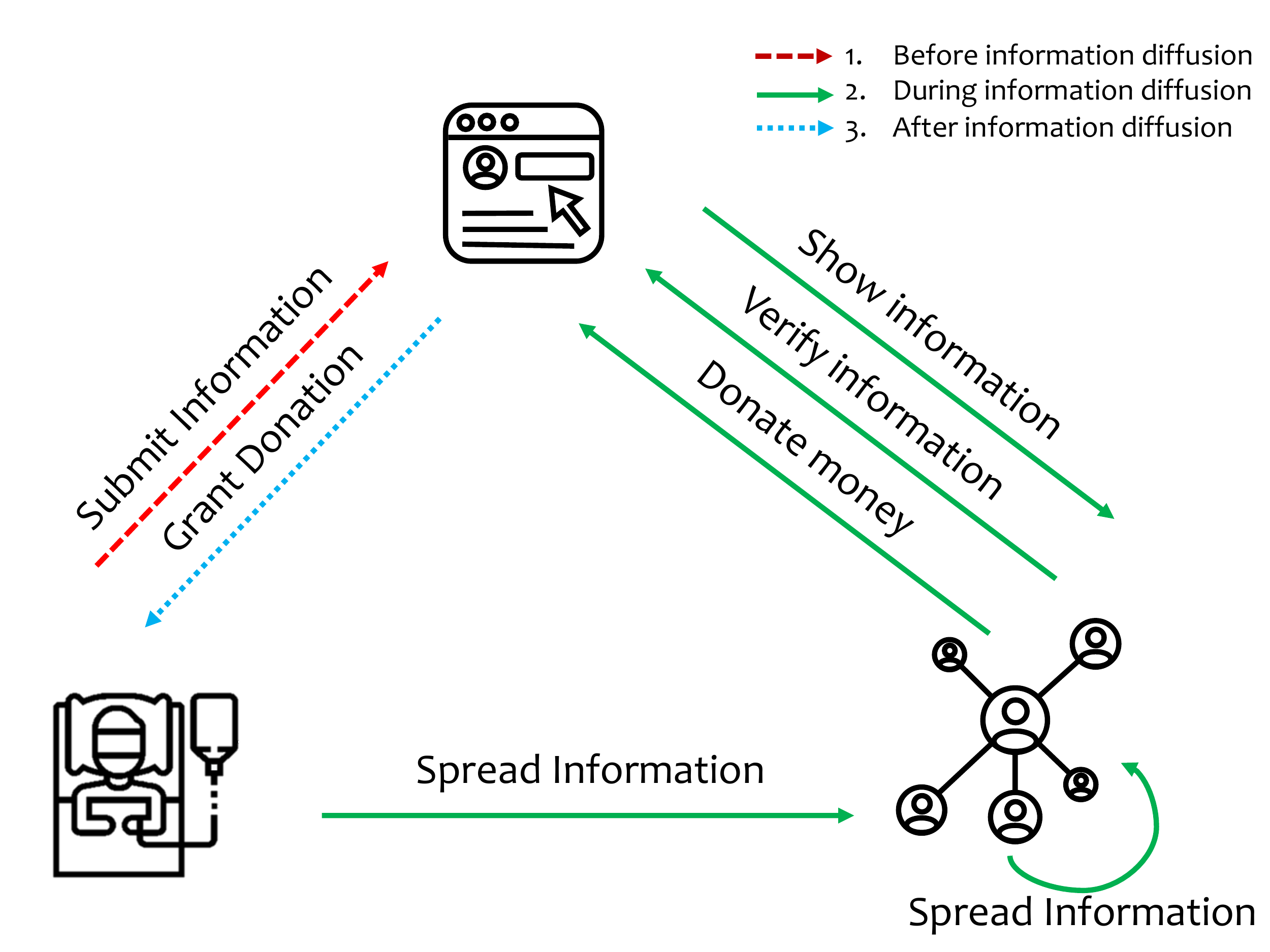}
  \caption{Medical fundraising on the web can be divided into three phases: (i) before the information diffusion, a fundraiser proposes a new case to the platform to conduct a preliminary review. (ii) during the information diffusion, users use their social networks to spread the crowdfunding information and call for donations. The platform collects verified information and donations from social networks. (iii) after the information diffusion, the platform grants the donations to the fundraiser.}
  \label{fig:shuidi-workflow}
\end{figure}

We illustrate how medical crowdfunding works in \figref{fig:shuidi-workflow}.
For online medical crowdfunding, there are usually three key components: \textbf{a fundraiser}, \textbf{a platform} and \textbf{the social network of the fundraiser}.
We divide a fundraising campaign into three parts: before information diffusion, during information and after information diffusion.
Before information diffusion, the fundraiser first submits some information on the platform, and the platform reviews the authenticity of the case.
After that, the fundraiser begins to spread his fundraising request to the social network (\eg the WeChat Moment or WeChat Group).
In addition to resharing and donating, people in the social network can also verify the fundraising cases to increase the credibility of the fundraising campaigns.
This verified information usually includes verifying the authenticity and expressing the relationship with the fundraiser. 
For example, \textit{This patient is my friend; he is recently suffering a severe disease. I hope everyone gives him a hand.}
People can easily continue these activities on social networks.
The platform collects information, and donations from social networks and strengthens the crowdfunding campaigns.
After the fundraising is completed or finished by the fundraiser, the platform will grant the collected money to fundraisers to treat the disease.
The fundraising model illustrated in \figref{fig:shuidi-workflow} is adopted by most  medical crowdfunding platforms in China.

Based on the forementioned fundraising model, we focus on answering two research questions in this paper:
\begin{enumerate}
\item Given the model in \figref{fig:shuidi-workflow}, \textbf{what factors affect fundraising campaigns at different phases of information diffusion?}
  More specifically, before the information diffusion, what social status factors will affect the fundraising campaign?
  During the information diffusion, what social network factors will affect the fundraising?
\item Based on social status factors and social network factors, can we build a predictive model that predicts future fundraising result in the early stages of information diffusion?
\end{enumerate}

\vpara{Organization}
The rest of the paper is organized as follows.
Firstly, we introduce the relevant background and related works.
Then, we describe our dataset and do data cleaning.
The dataset comes from a popular online fundraising platform in China, which spans a month in March of 2019, including a total of 27,000 cases.
We find a low achieved ratio in these fundraising campaigns.
Further, we quantitatively study the factors which affect the fundraising, including the impact of social status and social network.
Last but not least, we formulate donations prediction tasks using different machine learning methods.
We adopt some methods from popularity prediction and achieve good predicted results.
Finally, we offer our concluding thoughts and discuss future work.


\label{sec:1-intro}

\section{Background and Related Work}\label{sec:2-background}

In this section, we will briefly introduce some backgrounds and related works about medical crowdfunding.

\subsection{Crowdfunding Mechanism}

Crowdfunding is defined as an open call mostly through the Internet for the provision of financial resources by a group of individuals instead of professional parties either in the form of donations, in exchange for a future product, or exchange for some kind form of reward \citepp{belleflamme2014crowdfunding,schwienbacher2010crowdfunding,gerber2013crowdfunding}.
Crowdfunding typically contains three participating stakeholders: \textit{the project initiators} who seek funding for their projects, \textit{the backers} who are willing to back a specific project, and \textit{the matchmaking crowdfunding platforms} acting as intermediaries \citepp{gierczak2016crowdfunding,belleflamme2014crowdfunding}. 
In this paper, we refer to three participate stakeholders as \textbf{fundraisers}, \textbf{donors}, and \textbf{platforms}.
Many researchers have discussed and studied the relationship between the three parts, their motivations, benefits, and risks \citepp{gerber2013crowdfunding,haas2014empirical,belleflamme2014crowdfunding,gierczak2014all}.
It is worth mentioning that past research has mainly discussed commercial crowdfunding, such as Kickstart.
With the development of crowdfunding, non-profit organizations and government organizations have also used crowdfunding for social good (\eg alleviate poverty\footnote{https://www.zgshfp.com.cn/}). 
Medical crowdfunding is different from the other models mentioned above.
We compare these different crowdfunding models from  platform, target, risk, and time in \tableref{tab:difference}.

\begin{table*}[!htp]
  \centering
  \caption{The differences between medical crowdfunding, poverty alleviation crowdfunding, and commercial crowdfunding}
  \label{tab:difference}
  \scalebox{1.0}{
    \begin{tabular}{cccc}
      \toprule
      &  Medical crowdfunding   & Poverty alleviation crowdfunding & Commercial crowdfunding\\
      \midrule
      Target & Healing disease &  Poverty alleviation & Accomplishing plan\\
      Time & In a short time  & In a long time & In a longer time\\
      Platform Organizer & Companies & Goverments & Companies \\
      Reward & No reward & No reward & Reward based on donations\\
      Fundraiser & Patient & The poor & Organization/Individual\\
      Fee & No fees &  No fees & Some fees (\eg 5\%)\\
      Risk &  No risk & No risk & High risk\\
      Utility of information diffusion  & Important & General & General\\
      Relationship with fundraiser & Close relationship & Strange relationship & Based on interest\\
    \bottomrule
  \end{tabular}
  }
\end{table*}

From \tableref{tab:difference}, medical crowdfunding is primarily fundraising for treating personal diseases, which uses social networks to disseminate and verify the information. 
The utility of information diffusion is much higher than others.
It can be considered that social capital \citepp{lin2001social} has played an important role in crowdfunding for serious illness.
Based on this characteristic, we analyze the social factors that affect medical crowdfunding in this paper. 

\subsection{Social Factors in Crowdfunding Campaign}
For crowdfunding research on social computing, researchers would focus on some
social phenomenons or some social factors (\eg, motivations and deterrents \citepp{gerber2013crowdfunding}, geographical and organizational distances \citepp{muller2014geographical}, donor retention \citepp{althoff2015donor}, trust amid delays \citepp{kim2017understanding}, credibility factors \citepp{kim2016power} and so on) to analyze the crowdfunding mechanism.
However, the related crowdfunding studies are mainly about commercial crowdfunding, which aims at raising money for their own commercial ventures \citepp{tanaka2016legitimacy}.
For philanthropic crowdfunding, \citea{tanaka2016legitimacy} demonstrate that the legitimacy of a philanthropic crowdfunding campaign is a central concern for fundraisers.
For the medical crowdfunding, \citea{kim2016power} interviewed fifteen people involved in medical crowdfunding and found that support networks were larger than beneficiaries expected, with strangers offering support. 
Supporters offered not only monetary but also volunteering contributions, including campaign creation, promotion, and external support \citepp{kim2017not}.
These works are close to our findings in this paper, but they are mainly qualitative analysis. 
In this paper, we devote to finding what social factors will affect the success in medical crowdfunding based on mining real log datasets. 
Besides, the researches above mainly study on crowdfunding in the US, while social media platforms in different countries have different credibilities and characteristics \cite{yang2013microblog}.

\subsection{Crowdfunding Prediction}
%

The popularity of web content is the amount of attention that web content receives \citepp{gao2019taxonomy,tatar2014survey}. 
Users public and share the content with his friends, and several of these friends share it with their respective friends, developing a cascade of resharing can develop, potentially reaching a large number of people \citepp{cheng2014can}.
The particular types of popularity prediction involves tweet/microblogs \citepp{bao2015modeling,bao2013popularity}, images \citepp{zhang2018user}, videos \citepp{rizoiu2017expecting}, recipes \citepp{sanjo2017recipe}, academic papers \citepp{shen2014modeling} and so on.
According to the timing of prediction, the popularity prediction of web content in online social networks can be divided into two types: ex-ante prediction and early prediction \citepp{hofman2017prediction}.
The ex-ante prediction uses only information available before a given cascade.
The early prediction will observe the progression of a cascade for some time before making a prediction \citepp{hofman2017prediction}. 

Many researchers have applied machine learning to predict whether crowdfunding can be successful, but most of the research has focused on ex-ante prediction\citepp{cheng2019success,lee2018content}.
In the context of medical crowdfunding, information diffusion is an important mechanism, and widespread cases have more donation amounts and donate counts.
\citea{kindler2019early} find that some patterns at the very beginning of the campaign can be an excellent success predictor.
These findings are close to some studies on popularity prediction.
The studies on popularity prediction can help us better analyze, model, and predict medical crowdfunding donations.
We adopt some machine learning methods on popularity prediction to our prediction of donations tasks.
Our methods belong to early prediction, which usually has a better performance than ex-ante prediction.



\section{Dataset}
\label{sec:3-dataset}
Our dataset is provided by \shuidichou ~(ShuidiChou), which is one of the most popular online medical crowdfunding platforms in China. 
According to the reports\footnote{https://finance.yahoo.com/news/waterdrop-crowdfunding-topped-among-chinese-091500004.html},
from 2016 to 2019, Waterdrop Fundraising helped patients suffering from difficult and serious diseases to raise more than 20 billion yuan of medical funds in total. 
The number of donations has exceeded 500 million, and the monthly growth has been maintained at over 1.5 billion (CNY).

\subsection{Dataset Description}
Our dataset is randomly selected from logs spanning a month in March of 2019, including a total of 28,000 cases. 
According to the fundraising rules, the fundraising period for a case is 30 days, and the fundraiser can end the fundraising cycle within 30 days.
We removed some invalid cases whose fundraising periods are more than 30 days.
In total, there are 27,618 cases. 
The earliest start time for these cases was 00:04:38 on March 1, 2019, and the latest end time was 23:50:05 on March 31, 2019. 
These cases have a total of 14M shared counts, 710K verifications, 20M donated counts, and 700M (CNY) raised money in total.
With keeping data only necessary for our research,  our dataset has two parts: the activity log and the detailed case information.

The log includes users' shared, verified, and donated behavior.
The log of each shared activity includes shared user id (\eg $A$), user source id (\eg $B$), shared time (\eg $t^{s}_i$), shared case id (\eg $c_i$), and channel (\eg WeChat).
Especially, it means a user $A$ shared a case message $M_{c_i}$ using WeChat at time $t^{s}_i$.
And the message $M_{c_i}$ is forwarded from the user $B$.
The log of each donate activity involves donated user id (\eg $A$), donated case id (\eg $c_i$), donated amount (\eg 10 CNY), and donated time (\eg $t^{d}_i$).
It means a user $A$ donates 10 CNY at time $t^{d}_i$ for $c_i$.
The log of each verification includes verified user id (\eg $A$), verified time (\eg $t^{v}_i$), case id (\eg $c_i$), and relationship (\eg relatives).
It means a user $A$ verify case $c_i$ as the relative of the fundraiser at time $t^{v}_i$. 
Besides, after removing sensitive user information to protect personal privacy, we obtained detailed information for each case, including information about the fundraiser (age, gender, illness, domicile province, and hospital province), target amount, final obtained amount, and case-related text information (title and content).

\subsection{Inequality in Donations}
First, we investigated whether there is an inequality in medical crowdfunding on the web.
In online fundraising, the target amount (i.e., goal) is related to the fundraiser's expectations, reflecting the fundraiser's needs and estimates for fundraising before spreading the fundraising information.
The average fundraising goal is 172,250, and the average obtained amount is 25,732.1.
The average ratio of achievement is about 0.172, which is lower than the results in GoFundMe \citepp{berliner2017producing}.

\begin{figure}[!ht]
	\centering
	\includegraphics[width=0.5\textwidth]{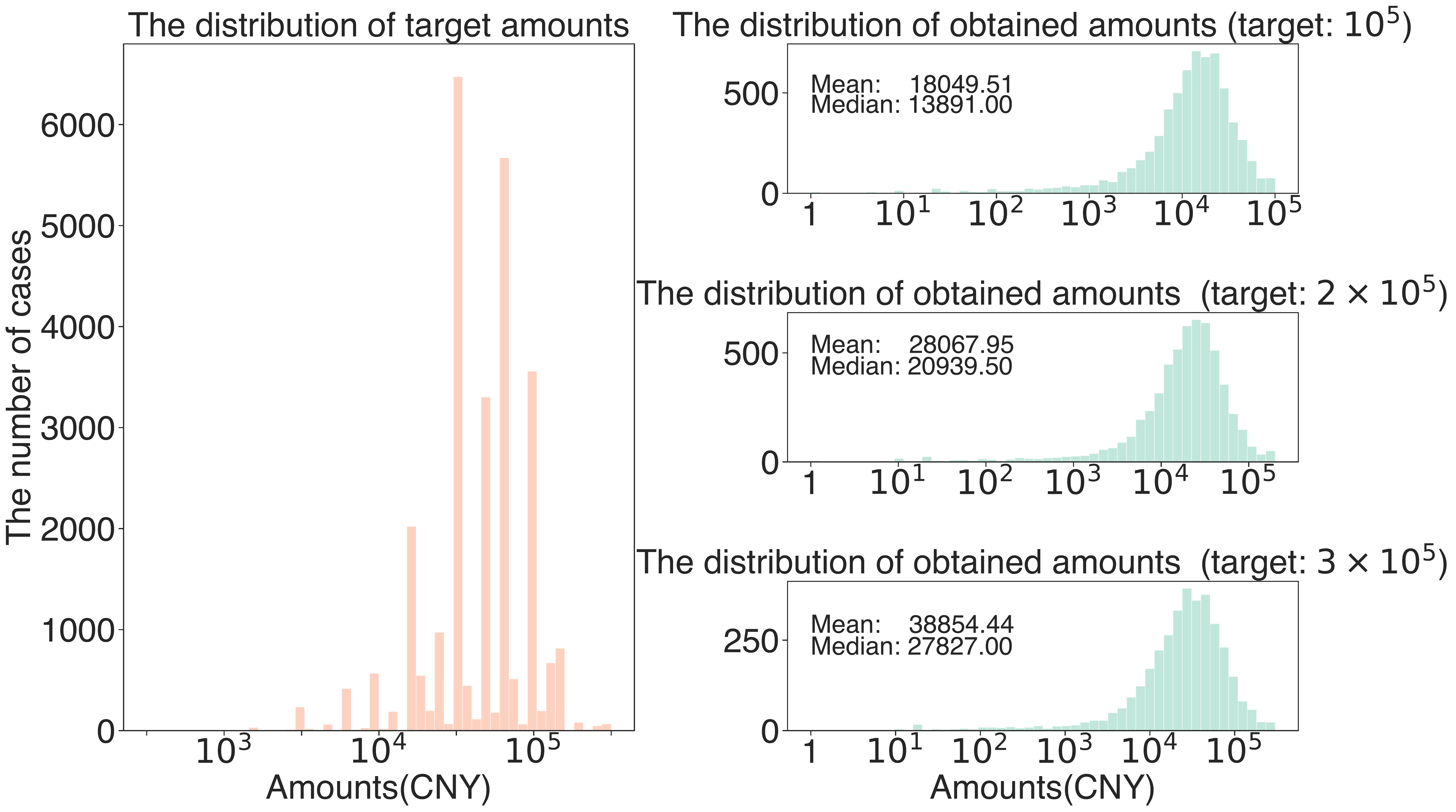}
	\caption{The distribution of fundraising target amounts and fundraising
    obtained amounts.}
	\label{fig:distribute-obtain-target}
\end{figure}
In \figref{fig:distribute-obtain-target}, we can find that the expectation of fundraising is mainly concentrated between 50K and 100K, but most of the fundraising cases can only raise about 10K.
In the same target, different people will raise different amounts.
Only very few people can achieve their fundraising goals.

\subsection{Factors Affecting Fundraising}

As we mentioned earlier, the achieved ratio is relatively low, indicating that few fundraisers can finish their fundraising goals.
In commercial crowdfunding, whether the goal can be achieved depends on the competitiveness of the project itself, such as commercial value and return.
However, in the context of medical crowdfunding, it mainly reflects the social capital of fundraisers.
For a medical fundraising campaign, there mainly exist two factors affecting the fundraising, \ie social status factor and social network factor.
On the one hand, some people may have a higher social status, higher educations, and higher income.
Some people may have more friends or their friends are more wealthy because of their social class.
These factors can be viewed as social status or social profile factors.
On the other hand, the social network plays a very vital role in the fundraising campaign \citepp{tanaka2016legitimacy}.
Specifically, some people can make better use of their social networks to spread information more widely, getting more people's help and raising more money.
In order to understand and analyze the factors affecting the success of fundraising, we discussed the role of social status and social network separately.


\section{The Impact of Social Status} 
\label{sec:4-phase1}
In this section, we focus on social status factors.
The platform usually requires the fundraiser to provide some necessary information like age, name, mobile phone, diagnosis sheet, and so on. 
Some information will be shown in the pages on the web, and others will be assessed by the platform.
Besides the mentioned necessary information, the fundraiser can edit the description content (\eg title and content) according to their situation.

We extract the information about the fundraiser: age, gender, target fundraising amount, domicile province, hospital province, number of diseases, and text-related features (\ie negative score, text content length, and the number of locations mentioned).
Notably, we use the Chinese word segment and sentiment analysis tools\footnote{https://intl.cloud.tencent.com/} to extract the negative text score.
Specifically, since the sentiment analysis tool has a limitation of the length of the analyzing text, we divide the text into chunks and then take the average sentiment score of the pieces as the final sentiment score.
The negative score represents the probability of negative emotions, and it ranges from 0 to 1.
If the value is less than 0.5, the text content is positive; otherwise it is negative.
Besides, we find that in the text, people usually will mention their hometown, residence, or workplace. 
Since that geographic information is an essential factor in fundraising, we count the number of the location mentioned in the content by matching a geographic dictionary. 
For extracting the number of diseases, we use some professional vocabulary to extract the disease categories\footnote{https://en.wikipedia.org/wiki/Lists\_of\_diseases}.
We correlate these factors with the amount of donations using Pearson Correlation Analysis.
To better show the impact of different factors on fundraising, we sort the cases according to the fundraising amount and calculate the feature statistics of the top 10\% cases and the last 10\% cases, marked as top10 and bottom10.
From \tableref{tb:factors-before-information-diffusion}, we can find:

\begin{figure}[!ht]
  \centering
  \includegraphics[width=0.5\textwidth]{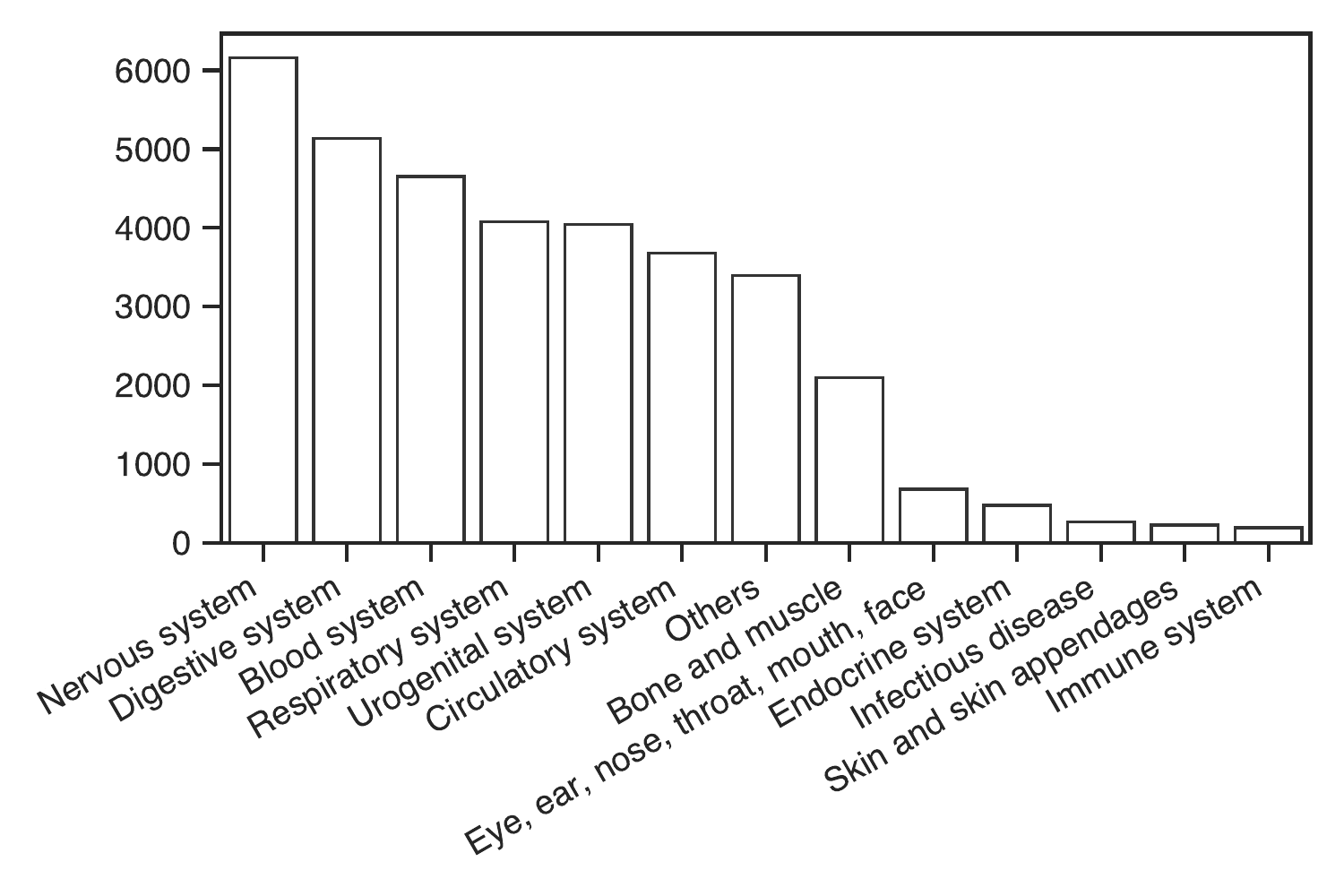}
  \caption{The barplot of disease classification.}
  \label{fig:disease-dist}
\end{figure}

\begin{table*}[!ht]
\begin{center}
\small
\caption{Analysis of the factors on the fundraising information before information diffusion (${}^{**}p<0.01$)}
\label{tb:factors-before-information-diffusion}
	\begin{tabular}{l r r  r r }
    \toprule
		Variable & Total & Top10\% & Bottom10\% & $R$ \\ 
		\midrule
    \multirow{2}{*}{
    Gender
    } & Male (61.3\%) & Male (61.5\%) & Male (61.1\%) & \multirow{2}{*}{$0.002$}\\
    & Female (38.7\%)  & Female (38.5\%) & Female (38.9\%) \\ 
		\midrule
		Age & 46.96 ($\pm$17.02) & 35.97($\pm$17.11) & 48.60 ($\pm$17.97) & $-0.155^{**}$\\
		\midrule 
		Target amount & $10^{5.131(\pm0.327)}$ & $10^{5.401(\pm0.220)}$ &  $10^{4.947(\pm0.426)}$  & $0.342^{**}$ \\
		\midrule
		Text content length & 445.70 ($\pm$225.62) & 572.15 ($\pm$286.73) & 358.36 ($\pm$209.28) & $0.228^{**}$\\ \midrule 
		Title content length & 20.18 ($\pm$4.59) & 20.47 ($\pm$4.45) & 19.44 ($\pm$5.34) & $0.055^{**}$\\ \midrule
		\# Diseases & 1.50 ($\pm$1.14) &1.58 ($\pm$1.12) & 1.03 ($\pm$1.05) & $0.145^{**}$ \\ \midrule 
		\# Locations mentioned (Province) & 1.77 ($\pm$0.89) & 1.90 ($\pm$0.97) & 1.62 ($\pm$0.93) & $0.084^{**}$ \\ \midrule 
		\# Locations mentioned (City) & 1.79 ($\pm$1.36) & 2.12 ($\pm$1.58) & 1.47 ($\pm$1.17) & $0.120^{**}$ \\ \midrule
		Negative score & 0.32 ($\pm$0.32) & 0.36 ($\pm$0.29) & 0.33 ($\pm$0.36) & $0.021^{**}$ \\ \bottomrule
	\end{tabular}
\end{center}
\end{table*}

\begin{itemize}
    \item In medical crowdfunding, although fundraisers are mostly male, gender is not an important factor for the amounts of donations.
    \item Younger fundraisers are more likely to raise more funds. This may be due to the fact that younger people are more familiar with the Internet and have more social capital. 
	\item The higher the target amount, the more funds will be raised. 
	This finding contradicts \citea{wash2013value}, which found that projects with lower
  fundraising goals tend to be more likely to reach their goal.
  The mechanisms in Wataerdrop platforms may cause this.
  Specifically, in \shuidichou, its fundraising rules will allow fundraisers to end fundraising without achieving their fundraising goal. 
	There are not any refunds, even if the fundraisers may not achieve their goals.
	So it is usually better for fundraisers to propose higher fundraising goals.
\item
  More detailed text descriptions, including more geographic location information (province and city), more text description, and more extended title, are positive related to the final fundraising amounts (e.g., the correlation coefficient R of text content length is 0.228).
  Besides, the fundraiser who raised more money than others is concentrated in economically developed areas. 
\item We plot the classification of fundraising in \figref{fig:disease-dist}.
  By analyzing the disease, we find that most fundraising diseases are cancers, such as lung cancer, liver cancer, etc.
  The proportion of external trauma and infectious disease is relatively low than internal organs. 
  This phenomenon is consistent with the positioning of ``Severe Illness'' instead of ``Sudden Illness''.
  These diseases are usually accompanied by the occurrence of other diseases (The number of diseases is large than one).
    \item Most of the fundraising content is positive (since the average negative score of total cases is 0.316, which is smaller than 0.5), which may be due to platform guidance. But more negative expressions will make it easier to arouse compassion and get more funds.
\end{itemize}

Generally speaking, the information submitted by the fundraiser will have a particular impact on the amount of fundraising. 
The fundraiser should prepare more detailed information about his situation (\eg more location referred) to make the people know your difficulty and offer their help.


\section{The Impact of Social Network}
\label{sec:5-phase2}
In this section, we analyze the role of social network mechanisms in the process of information diffusion, including the social utility of the platform, social network verification mechanism, and close neighbors in the information diffusion cascades.

\subsection{Social Utility of the Platform}

As we said before, in a case fundraising process, the platform participates in this fundraising information diffusion by showing the information on the platform. 
About 15\% of the shared messages come from the platform website.
It means that some users know and help the fundraisers by browsing the website of the platform (\eg Love Homepage, which shows some cases on the homepage).
Besides, these users also indirectly affect their neighboring users by sharing the information in their social networks.

\begin{figure}[!ht]
	\centering
	\includegraphics[width=0.45\textwidth]{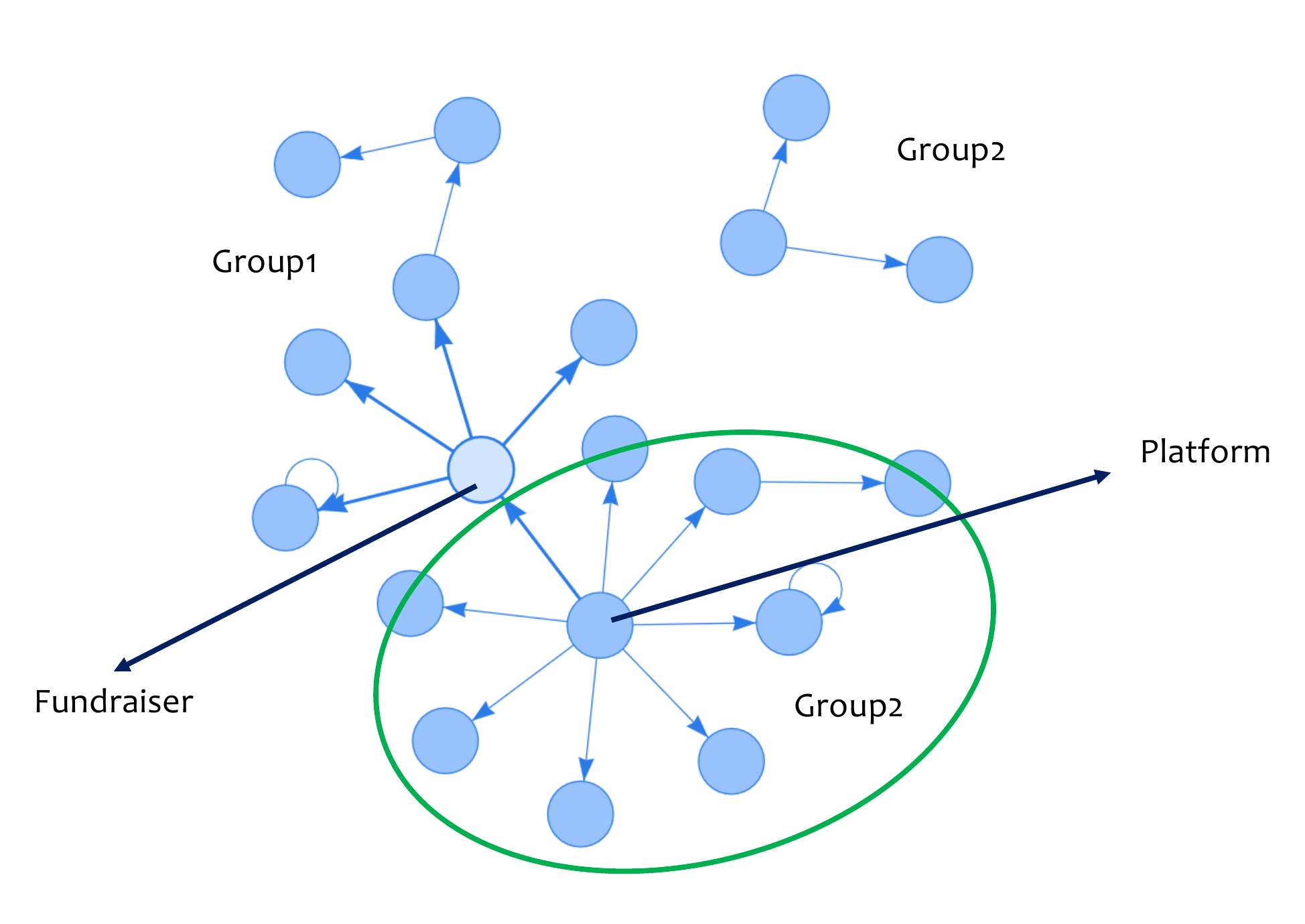}
	\caption{Illustration of our group partition. We divided the nodes which starting with the fundraiser as Group1 $G_1$ and others as Group2 $G_2$. $G_2$ is regarded as the social influence caused by the platform.}
	\label{fig:group-partition}
\end{figure}

\begin{figure}[!ht]
	\centering
	\begin{subfigure}[t]{0.23\textwidth}
		\includegraphics[width=\textwidth]{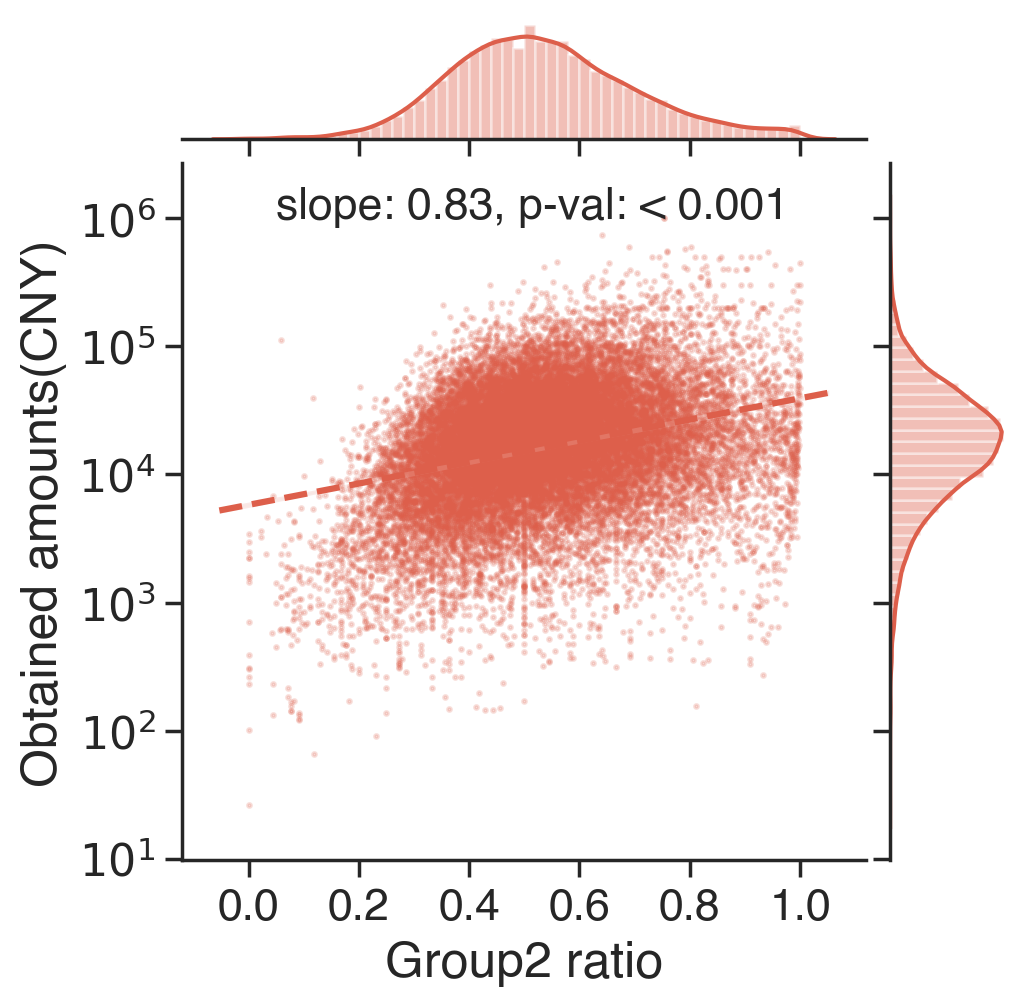}
		\caption{Group2 ratio and donated amounts.}
		\label{fig:surrender vs winrate}
	\end{subfigure}
	\begin{subfigure}[t]{0.23\textwidth}
		\includegraphics[width=\textwidth]{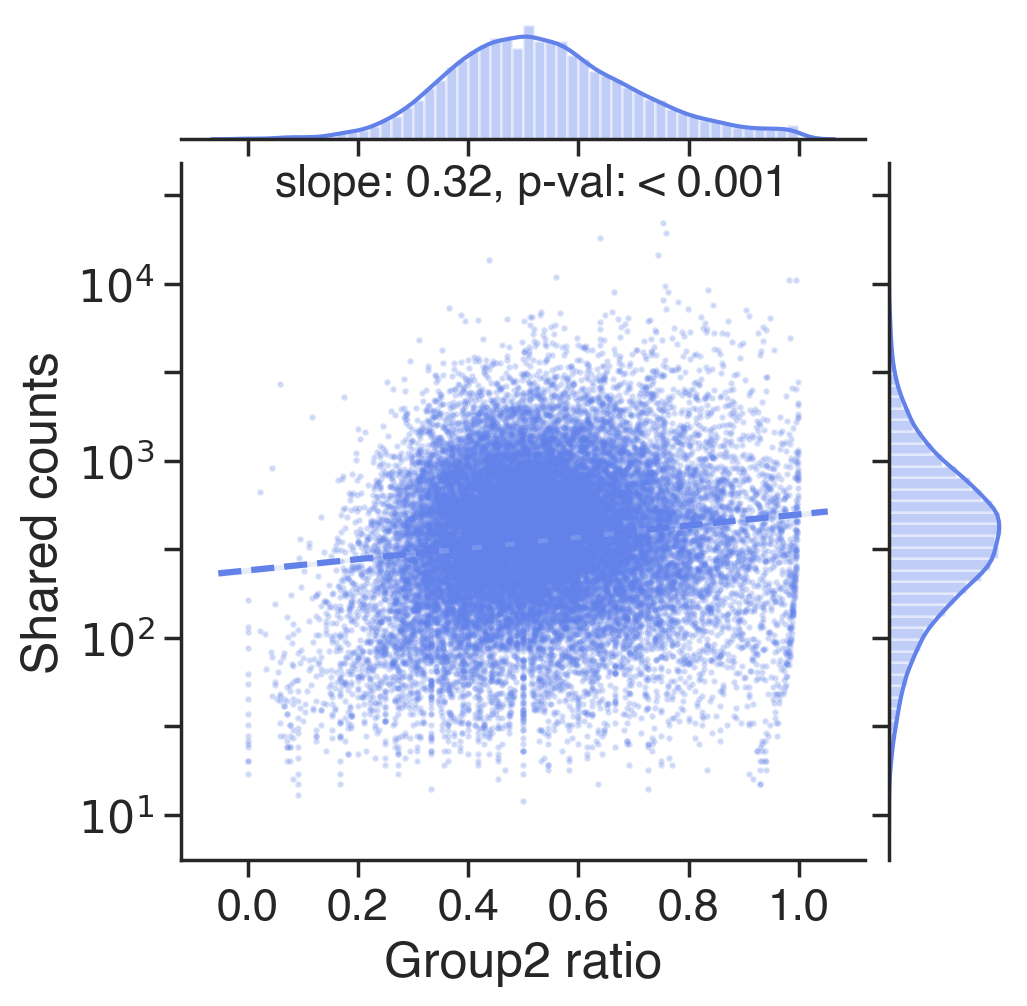}
		\caption{Group2 ratio and shared counts.}
	\end{subfigure}
	\caption{The relationship between $P_{G_2}$ and obtained amounts, shared counts.}
	\label{fig:group2-ratio}
\end{figure}

To measure the role of the platform in the information diffusion, we build the diffusion cascade graph in action time order (\eg if B shared a fundraising message from A, we have a link from A to B. These links are generated in time order).
The starting node for fundraising is the platform (\ie the first link is from platform to fundraiser, which is that fundraiser share his message from the platform.). 
As shown in \figref{fig:group-partition}, we mark the nodes which starting with the fundraiser or fundraiser's successors as Group1 $G_1$. 
We think this group can be  social network of fundraiser can use without platforms.
The nodes that cannot be started from the fundraiser are marked as Group2 $G_2$, which is regarded as the impact of platforms.
We calculate the proportion of $G_2$ as $P_{G_2} = \frac{|G_2|}{|G_1| + |G_2|}$. 
It reflects the social utility of the platform in different cases.

We analyze the relationship between ratio of platform $P_{G_2}$ and the number of shares and donations in \figref{fig:group2-ratio}.
Note that, the larger the proportion $P_{G_2}$, the more users come from the platform's propagation utility.
Although some user source ids are missing and $G_1$ may be slightly different from the ground truth, $P_{G_2}$ in all cases can still be an approximate measure of the social utility of platform in information diffusion.

From \figref{fig:group2-ratio}, we can find that the proportion $P_{G_2}$ shows a positive correlation to the number of shared counts and obtained amounts. 
In other words, the platform will benefit information diffusion and donations. 
If someone just raises funds himself without using the platform's diffusion utility, it will be challenging to get people's attention and donations.
The platform, as part of the fundraising on the web mechanism, has enlarged the influence and spread of fundraisers.

\subsection{Social Network Verification}
\begin{figure}[!ht]
	\centering
	\includegraphics[width=0.4\textwidth]{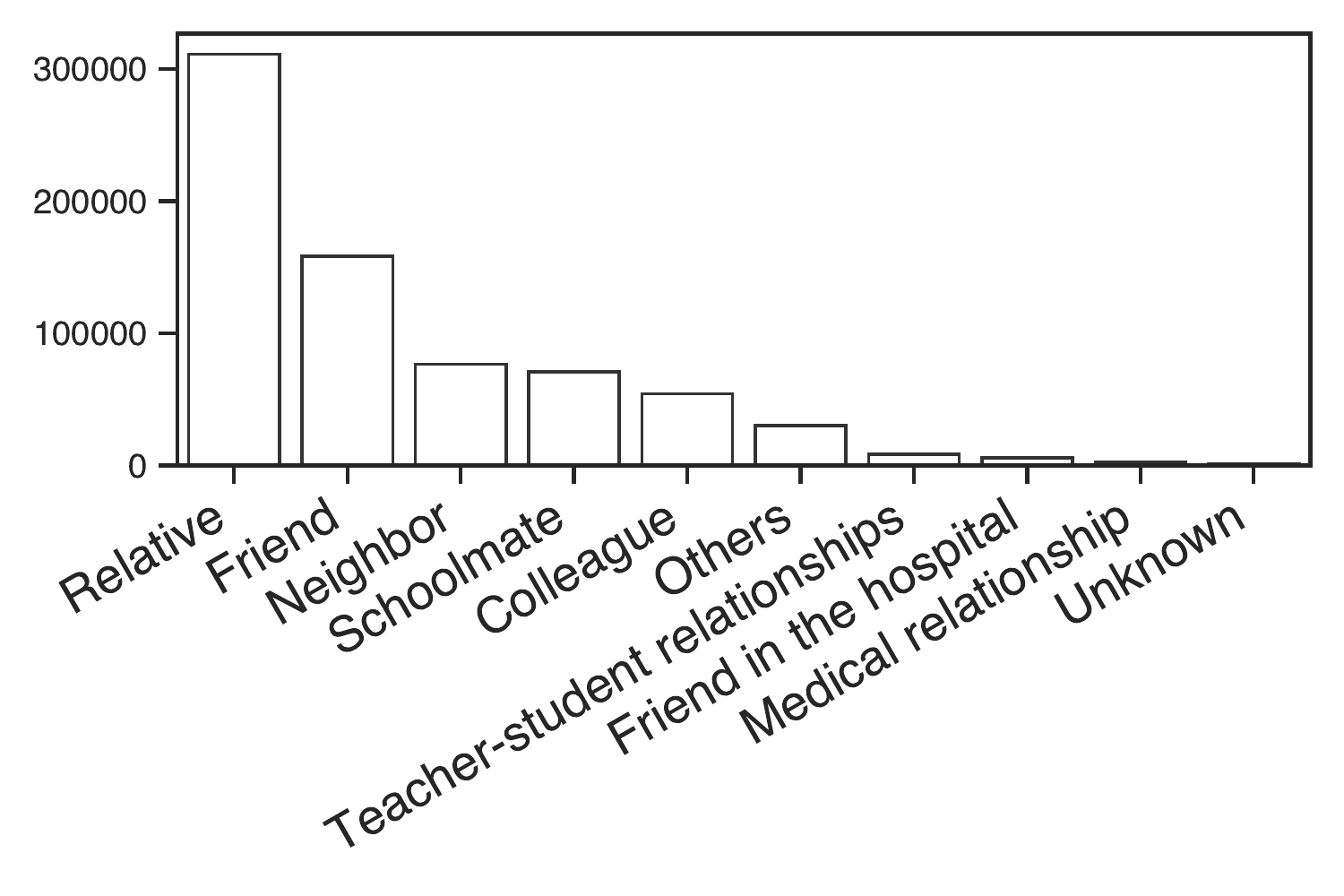}
	\caption{The barplot of the relationships between verified users and fundraisers.}
	\label{fig:verify-users}
\end{figure}

During the process of the fundraiser spreads the case on his social network, he will also look for someone to confirm his fundraising case. 
It's called social network verification. 
Those verified people usually confirm that the fundraising campaign is authentic and demonstrate the relationship with the fundraiser.
We counted the relationship between these verified users and fundraisers. 
From \figref{fig:verify-users}, we can see that the primary verified relationships are relatives and friends.
These relationships are strong ties instead of weak ties \citepp{granovetter1977strength}, indicating that people always try to find some close friends to verify their cases.

\begin{figure}[!ht]
  \centering
	\begin{subfigure}[t]{0.23\textwidth}
		\includegraphics[width=\textwidth]{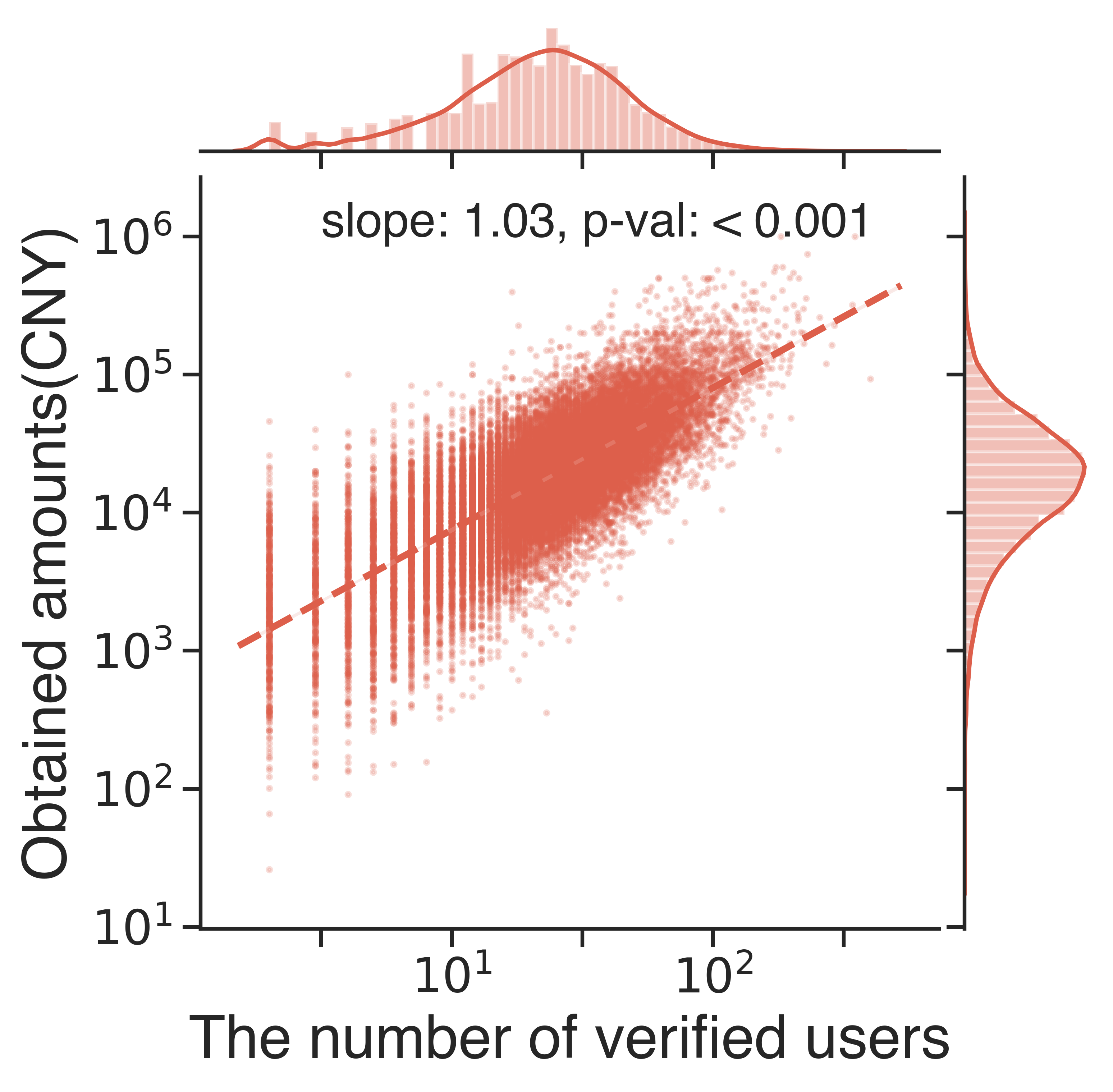}
		\caption{Verifed users and obtained amounts}
	\end{subfigure}
	\begin{subfigure}[t]{0.23\textwidth}
		\includegraphics[width=\textwidth]{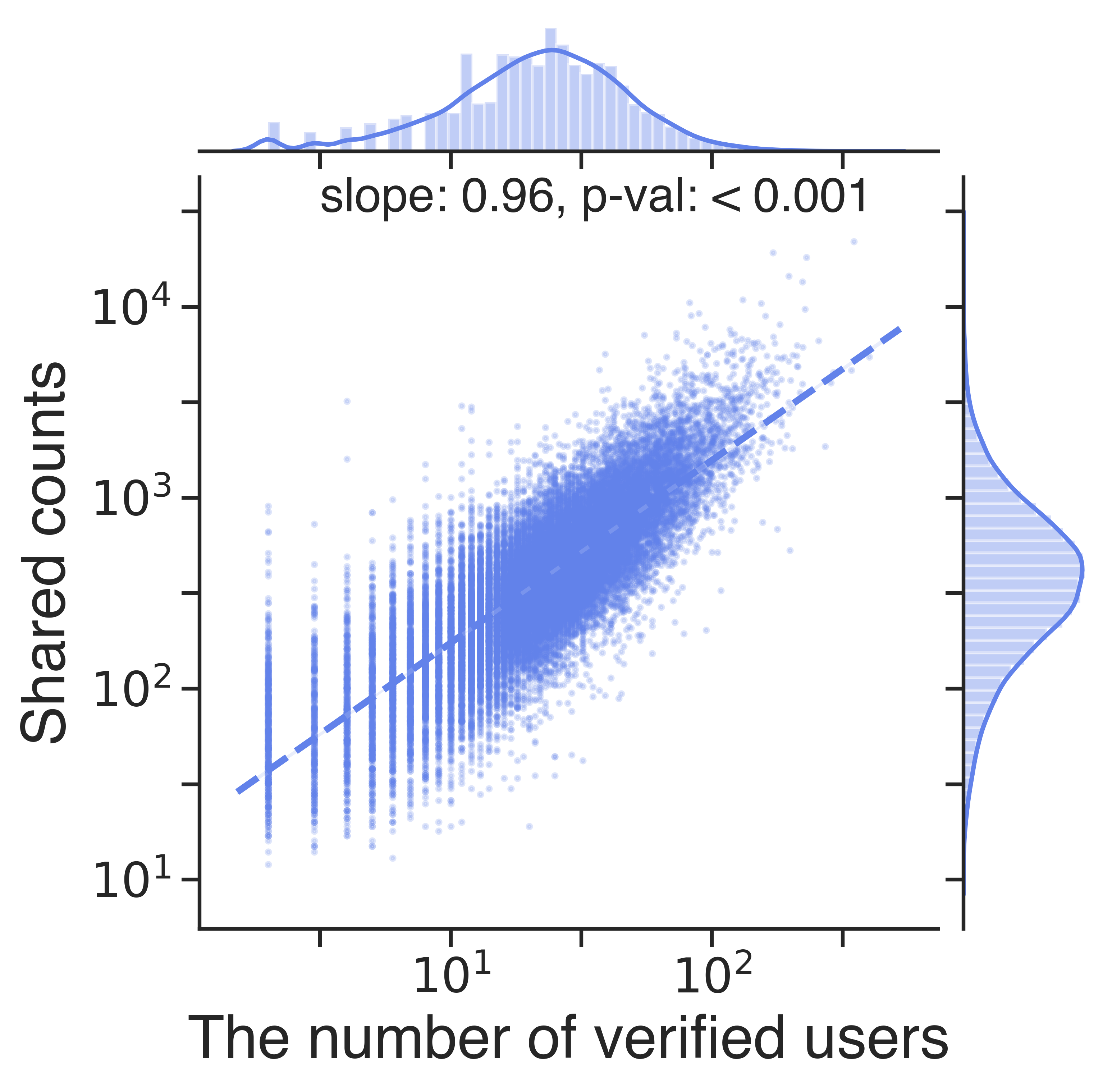}
		\caption{Verifed users and shared counts}
	\end{subfigure}
	\caption{The relationship between the number of verified users and obtained amounts, shared counts.}
	\label{fig:verify-information}

\end{figure}

We count the verified users and correlate it with the number of shared counts and obtained amounts.
\figref{fig:verify-information} shows that with more verified users, the amount of fundraising increased. 
In particular, when the number of verified users is higher than 100, the fundraising amount is usually more than 10,000, and the number of people sharing is usually higher than 1,000. 
The number of verified people may help the credibility of the fundraising case and further enhance the effectiveness of the information diffusion.
Compared with donations and reposts, it has proven to be another important type
of support for fundraisers in the medical crowdfunding \citepp{kim2017not}. 

\subsection{Donation in the Information Cascade}

\begin{figure}[!ht]
	\includegraphics[width=0.5\textwidth]{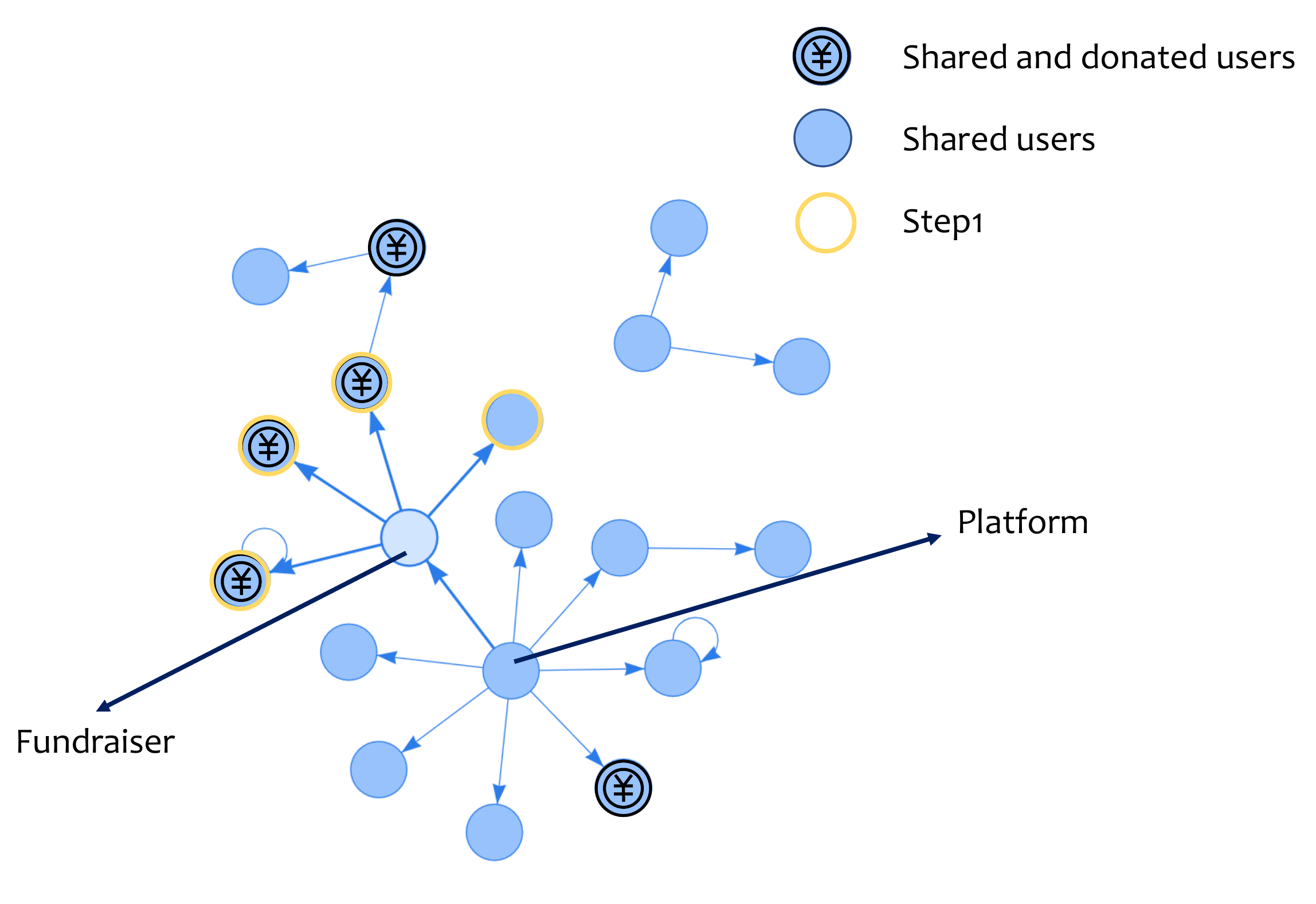}
	\caption{We define users who share the fundraising case from the fundraiser as the `step1'. We define other users as `other'. In \figref{fig:donation-partition}, the proportion of donation for `step1' is $\frac{1}{2}$ }
	\label{fig:donation-partition}
\end{figure}

In this section, we focus on the proportion of donations and the average amount of donations for users who have different closeness with fundraisers in the social network.
We define the users who directly share the fundraising case from the fundraiser as the `step1' while defining other users as `other'.
The `step1' is regarded to have a  more strong tie with the fundraiser than `other'. 
In WeChat, people can directly share the information of the fundraiser (\ie shared source id is the fundraiser) only if they are friends with the fundraiser.
We calculate the proportion of donation and average donation amount for `step1' and `other' in the information cascade.
We find that the proportion of donation in `step1'(0.647) is significantly lower than `other'(0.679), which p-Val $<$ 0.001. 
But the average donation(101.9) is significantly higher than `other'(60.5), which p-Val $<$ 0.001.
Generally speaking, the users of `step1' are close to the fundraisers, but their donation proportion is lower than `others'.
According to Chinese culture, one possible explanations is that people with close relationship are inclined to provide help to fundraisers offline, making the proportion of donations is relatively low.
Meanwhile, because of the location in the social network, `step1' users have a more intimate relationship than the other with the fundraiser, resulting in a large donate amount.
As for those users that are far away from the fundraiser, they usually give a small amount of fundraising conservatively.
This may be due to that, for the help of the friend of the friend, `other' users will participate in the repost and donate, but the weak tie will lead to a small donation amount.

\section{Prediction of Donations}
\label{sec:6-experiments}
\begin{figure}[t]
  \centering   
  \includegraphics[width=0.5\textwidth]{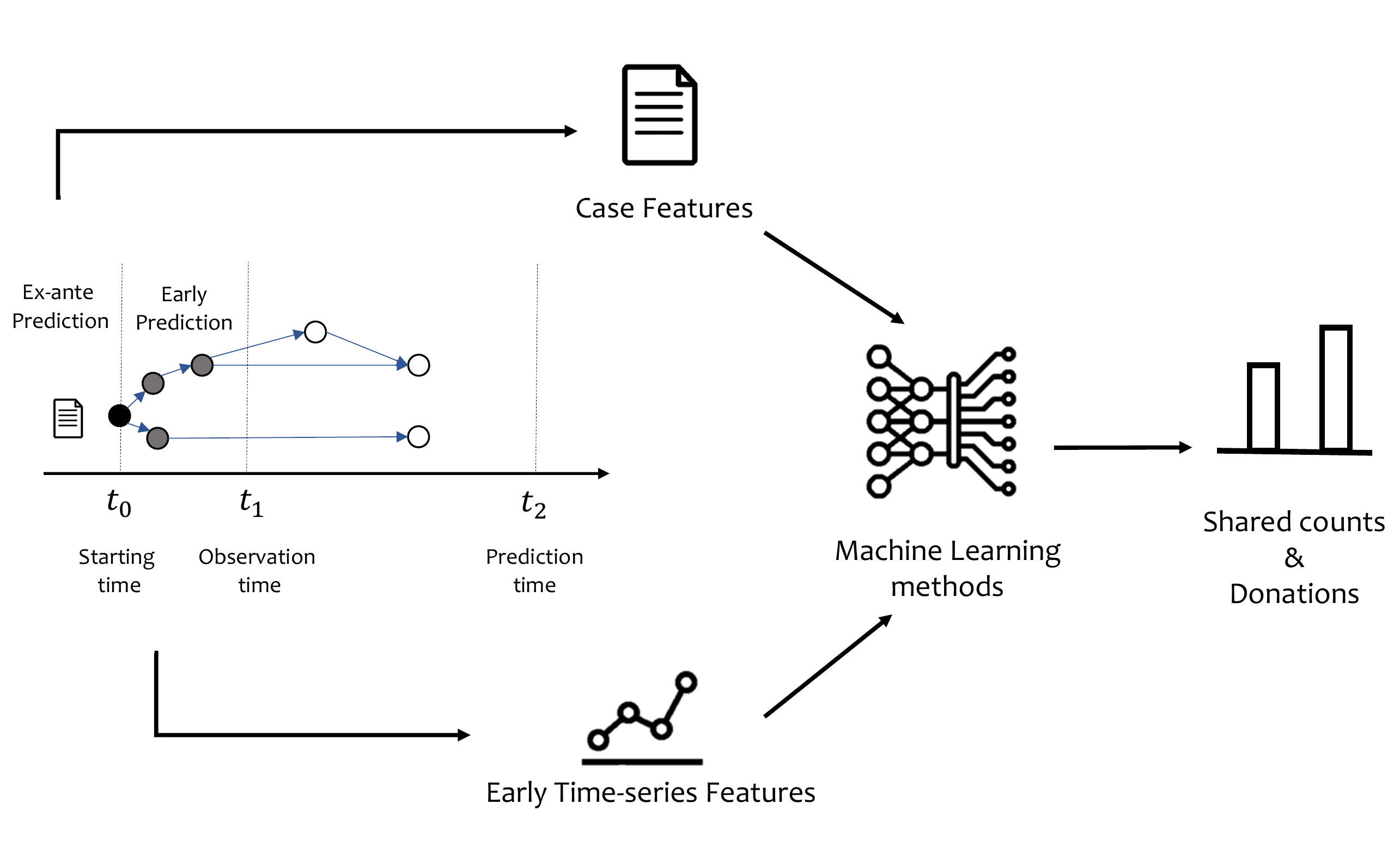}
  \caption{The black node is the original fundraiser, the gray nodes represent the shared users in the observation window, and white nodes represent unobserved ones. We select case features and early temporal features and use machine learning methods to predict the amount of sharing and donating.}
  \label{fig:machine-learning-methods}
\end{figure}
The popularity of web content is the amount of attention that web content receives, which can be measured in many ways such as the number of views or the number of `likes' \citepp{gao2019taxonomy}. 
In medical crowdfunding, we focus on the number of shared activities and the total amount of donations like \figref{fig:machine-learning-methods}.
In social status impact, we find that the age of the fundraiser, illness, case description, and other factors are related to the total donation amounts before the information diffusion. 
These case features can be viewed as ex-ante features \citepp{gao2019taxonomy}.
In the process of information diffusion, the early temporal patterns are also related to the final shared counts and total donations.
We can use some early observations to predict future popularity.

Here, we define our \textbf{prediction of donations} as following:
For a medical crowdfunding campaign, given the ex-ante features $X_{content}$ and the early temporal features $X_{t}$ in the
observation time window $[0, T]$, it predicts the shared counts $y^{s}$ and the donated amounts $y^{d}$.
For the case features $X_{content}$, we use the features mentioned before, including age, target amount, text content length, and so on.
For the temporal features, we extracted the increments of the three actions (\ie donate $y^d_{\Delta t}$, share $y^s_{\Delta t}$, and verify $y_{\Delta t}^v$) in the time interval $\Delta t$ (For donate actions, it includes donate counts $y^{cnt}_{\Delta t}$ and donate amounts $y^{d}_{\Delta t}$), which is 
\begin{equation}
  \begin{aligned}
    & X_{t} = (X^{s}_{t}, X^{d}_{t}, X^{d_{cnt}}_{t}, X^{v}_{t}), \\
    & X^{s}_{t} = (y^{s}_{\Delta t}, y^{s}_{2 \Delta t} -y^{s}_{\Delta t}, ...,  y^{s}_{T} -y^{s}_{T - \Delta t}) ^ T , \\
    & X^{d}_{t} = (y^{d}_{\Delta t}, y^{d}_{2 \Delta t} -y^{d}_{\Delta t}, ...,  y^{d}_{T} -y^{d}_{T - \Delta t})^T,  \\
    & X^{d_{cnt}}_{t} = (y^{d_{cnt}}_{\Delta t}, y^{d_{cnt}}_{2 \Delta t} -y^{d_{cnt}}_{\Delta t}, ...,  y^{d_{cnt}}_{T} -y^{d_{cnt}}_{T - \Delta t})^T,  \\
    & X^{v}_{t} = (y^{v}_{\Delta t}, y^{v}_{2 \Delta t} -y^{v}_{\Delta t}, ...,  y^{v}_{T} -y^{v}_{T - \Delta t})^T .
  \end{aligned} 
\end{equation}

In this paper, we set the time interval $\Delta t$ to 1 hour. 
We sort all cascades by their publication time and take the first 80\% as a training set, the middle 10\% as a validation set, and the last 10\% as a test set.
For the length $T$ of the observation time window, we consider three settings, \ie $T$ = 1 day, 2 days, and 3 days. 
We choose relative square error (RSE) as our loss function, \ie $f_{loss}(\hat{y}, y) = (\frac{\hat{y}}{y} - 1)^2$, and mean RSE (mRSE) over all test data as the evaluation metric, which is used by \citea{cao2017predicting}.

With these features, we choose typical machine learning methods to predict the donations.
\begin{itemize}
  \item ANN($X_{content}$): In this model, the only case features are used to train two-layer neural networks with rectified linear unit (ReLU):
  \begin{equation}
    \begin{aligned}
      & \hat{y}^{s} = \mathrm{ANN_{s}}(X_{content}), \hat{y}^{d} = \mathrm{ANN_{t}}(X_{content}), 
    \end{aligned}
  \end{equation}
  where $\mathrm{ANN}$ is the two-layer neural networks.
  \item SH($X_{t}$) \citepp{szabo2010predicting}: This model is based on the observation that the future popularity of content is linearly correlated with its early popularity. The results are predicted by:
  \begin{equation}
    \begin{aligned}
      & \hat{y}^{s}= \alpha^{s} y^{s}_t + b^{s},
      \hat{y}^{d} = \alpha^{d} y^{d}_t + b^{d} 
    \end{aligned}
  \end{equation}  
  where $y^{s}_t = \sum X^{s}_t, y^{d}_t = \sum X_t^{d}$, $\alpha, b$ are the parameters of this model.
  \item ML($X_{t}$) \citepp{pinto2013using}: It is an extension of SH model, replacing the single predictor, \ie the cumulative popularity $y_t$ , with multiple increments in equal-size time interval during the observation time window. 
  The results are predicted by:
  \begin{equation}
    \begin{aligned}
    & \hat{y}^{s}  = \Theta^s X_t^{s} + b^{s}, 
    & \hat{y}^{d}  = \Theta^d X_t^{d} + b^{d}
    \end{aligned}
  \end{equation}
  where $\Theta, b$ are the parameters.
\item ANN($X_{t}$): We use time series features of all actions and replace the linear regression with two-layer neural networks with ReLU.
  \begin{equation}
    \begin{aligned}
      & \hat{y}^{s} = \mathrm{ANN_{s}}(X_{t}),
      \hat{y}^{d} = \mathrm{ANN_{t}}(X_{t}) 
    \end{aligned}
  \end{equation}
  where $\mathrm{ANN}$ is the two-layer neural networks.
  \item ANN($X_{t}$ + $X_{content}$): We add the case features into the neural networks.
\end{itemize}
These models are implemented using Pytorch\citepp{paszke2019pytorch} with the Adam optimizer ($\mathrm{Learning\ Rate}=0.01, \mathrm{Weight\ Decay}=1e^{-4}$).
We train 200 epochs and select the model that performs best on the validation set for test.

\begin{table}[t]
  \centering
  \caption{Prediction of donations in Waterdrop Fundraising (mRSE, lower is better). We have bolded the highest value of each column and underlined the second value.}
    \label{tb:popularity-results}
    \scalebox{0.7}{
    \begin{tabular}{c|c|c|c|c|c|c}
      \toprule
       Prediction &  \multicolumn{3}{c| }{Shared Counts}& \multicolumn{3}{c}{Donated Amounts}\\
       \midrule
       Obeservation Time & 1 day & 2 day & 3 day & 1 day
       & 2 day & 3 day \\
      \midrule
      ANN($X_{content}$)  & \multicolumn{3}{c|}{0.9289}& \multicolumn{3}{c}{0.9925} \\
      SH($X_{t}$)      &0.1414     &0.0694 &0.0436      &0.1254 &0.0058      &0.0338\\
      ML($X_{t}$)       &0.1247 & 0.0636& 0.0446& 0.1136& 0.0533& 0.0363\\
      ANN($X_{t}$) & \underline{0.1246} & \textbf{0.0583} & \textbf{0.0383} & \underline{0.1134} & \underline{0.0493} & \underline{0.0308}\\
      ANN($X_{t}$ + $X_{content}$) & \textbf{0.1218} & \underline{0.0584}& \underline{0.0388}& \textbf{0.1110}& \textbf{0.0489}& \textbf{0.0307} \\ \hline
  \end{tabular}
  }
  \end{table}
\tableref{tb:popularity-results} shows that only using case content will not work well for the prediction of donations and shared counts. 
In contrast, the temporal features will get a good result in the task, which is consistent with previous researches \citepp{gao2019taxonomy}.
Longer observation windows have better results, which makes the prediction easier.
ANN$(X_t+X_{content})$ achieves almost the best results in the prediction tasks.
Combining both temporal features and content features are better than only using temporal features in most settings.

These experiments give us a new perspective to analyze medical crowdfunding campaigns, which uses some machine learning methods to model and predict the campaign process.
Our model can be used as a new tool for the fundraisers and platform managers to monitor their crowdfunding campaigns.
They can make earlier decisions that are conducive to fundraising based on the predictions of the model, to raise funds for treatment better.
And these features can also better assist platform managers in designing better fundraising platforms and giving more valuable fundraising strategy suggestions.



\section{Conclusions and Future Work}
\label{sec:7-conclusion}
Taken together, we empirically study the effect of different phrases in medical crowdfunding campaigns.
First, we qualitatively compare the differences between different crowdfunding models, finding that information diffusion plays a vital role in medical crowdfunding when compared with traditional crowdfunding. 
Further, we find most people can't finish their goals.
We believe that information diffusion plays an important role in medical crowdfunding, comparing traditional crowdfunding.
We discuss what social factors will affect crowdfunding activity in different phases.
For social status impact, we find that some personal information affect the fundraising campaign, including demographic features and text content features. 
For social network impact, we analyze the utility of the platform, the social network verification mechanism, and the one-hop neighborhood.
Last but not least, we adopted some popularity prediction methods for predicting the shared counts and donation, which achieve a good result in the tasks.

We hope our research can shield light on the medical crowdfunding, help people to raise money to save their lives, and help the organizers to improve their platform.
For fundraisers, we can guide the fundraisers to describe their situation, including their geographic location information and detailed information to improve credibility.
They should ask more friends to verify their cases and call for more sharing.
For platforms, they can develop some monitor systems to make more early decision to promote the campaigns.
Their social network verification mechanism can be helpful for fundraising campaigns.
Besides, the utility of the platform is important for medical crowdfunding.
The platform should add user privacy, user fairness, and social responsibilities to the design of the platform.

As future work, we will try to study the effect of different cultures and different platforms in medical crowdfunding.
Different environments and cultures will have different influences on the medical crowdfunding mechanism, and these influences objectively change the design of the platform.
Comparing different medical crowdfunding in different countries is valuable for fundraising mechanism design.


\section{Acknowledgements}
We would like to thank Waterdrop for sharing the data, Dongyang Zhang for data processing and the anonymous reviewers for their helpful comments. This work is funded by the National Natural Science Foundation of China under Grant Nos. 62041207 and 91746301. This work is supported by Beijing Academy of Artificial Intelligence (BAAI). Huawei Shen is also funded by K.C. Wong Education Foundation.

\bibliographystyle{aaai21}
\bibliography{refs}
\end{document}